\documentclass[twocolumn,showpacs,showkeys,preprintnumbers,amsmath,amssymb]{revtex4}

\usepackage{graphicx}
\usepackage{dcolumn}
\usepackage{bm}

\begin{document}

\title{One-dimensional Hubbard model at quarter filling on periodic potentials}

\author{C.~Schuster and U.~Schwingenschl\"ogl}
\affiliation{Institut f\"ur Physik, Universit\"at Augsburg, 86135 Augsburg, Germany}

\date{\today}

\begin{abstract}
Using the Hubbard chain at quarter filling as a model system, we study the ground state properties of highly
doped antiferromagnets. 
In particular, the Hubbard chain at quarter filling is unstable against $2k_F$- and $4k_F$-periodic 
potentials, leading to a large variety of charge and spin ordered ground states.
Employing the density matrix renormalization group method, we compare the energy gain of the ground state
induced by different periodic potentials. For interacting systems the lowest energy is found for a
$2k_F$-periodic magnetic field, resulting in a band insulator with spin gap.  
For strong interaction, the $4k_F$-periodic potential leads to a half-filled Heisenberg chain  
and thus to a Mott insulating state without spin gap. This ground state
is more stable than the band insulating state caused by any non-magnetic $2k_F$-periodic potential. 
Adding more electrons, a cluster-like ordering is preferred.
\end{abstract}


\maketitle
\section{Hole doped antiferromagnets}
The study of low-dimensional doped antiferromagnets was triggered by investigations of 
the phase diagram of high temperature superconductors. Upon doping, they show a transition
from an antiferromagnetic to a superconducting state. On the other hand, the
investigation of one-dimensional compounds is also an active field. 
A simple form of low doping in a Cu-chain is the replacement of a few Cu ions by non-magnetic ions like Zn.
In this case, the system is well described by the Heisenberg model with some spins removed \cite{Eggert}. 
The charge and spin order of highly doped compounds, on the contrary, are not understood by far. 
A prototypical material revealing strongly doping dependent charge  order  and antiferromagnetism in a less than
half-filled band is the spin chain compound  (Ca,Sr)$_{14}$Cu$_{24}$O$_{41}$.
This composite crystal contains two different structural components along the $c$-axis: Cu$_2$O$_3$ ladders
and CuO$_2$ chains. Whereas the ladders contain only Cu$^{2+}$-ions, 
nearly all Cu$^{3+}$-ions are located on the chains \cite{Nuecker00}. 
Substitution of calcium on the strontium sites leads to a transfer of  holes from the chains into the ladders.
For the Ca-rich compounds one generally assumes one hole per
formula unit on the ladders and five on the chains.
Near the quarter-filled band antiferromagnetic order is established \cite{Nagata}.

\begin{figure}
\includegraphics[width=0.5\textwidth]{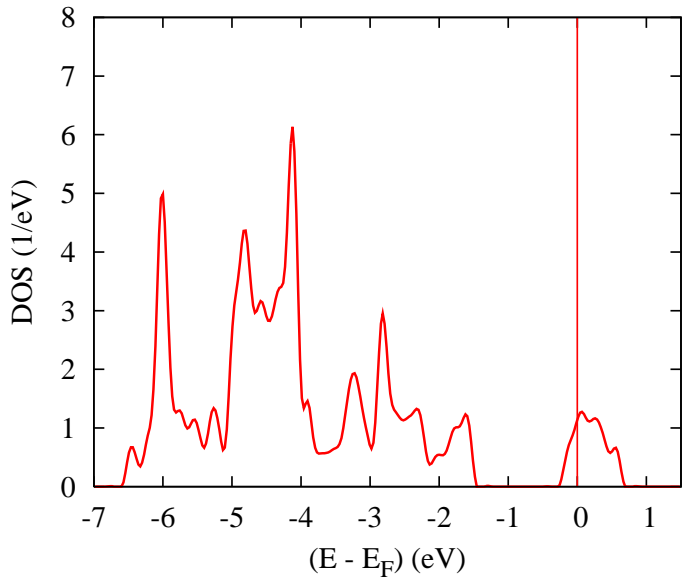}
\caption{(Color online) Partial Cu 3d density of states for the CuO$_2$-chains of Sr$_{14}$Cu$_{24}$O$_{41}$.}
\label{fig2}
\end{figure}

DFT-LDA calculations \cite{schwinud}, using the ASW method \cite{ASW}, reveal that the ladder and
chain substructures can be treated individually. In particular, the density of states of the chains
shows a partially filled valence band, as illustrated in Fig \ref{fig2}.
The origin of antiferromagnetism in these compounds is still unclear, 
and the doping dependence of the charge order and spin gap  is discussed controversially \cite{Klingeler}. 
Recent studies  on the basis of the Heisenberg model show that a simple spin model 
is not appropriate to describe  the magnetization 
\cite{Schnack}, indicating the necessity of studying more advanced models, as e.~g.\ the Hubbard model.

In this article we address the stability of the proposed charge order (cluster-like versus staggered)
\cite{Kataev} in (Ca,Sr)$_{14}$Cu$_{24}$O$_{41}$, considering  the competition between
kinetic energy, interaction and periodic distortions, on the basis of the Hubbard model. 
Since no periodic lattice distortion -- neither with two-site nor with four-site periodicity --
is found in the crystal structure \cite{Gotoh} we consider only periodic potentials.
Nevertheless, our results are relevant for all correlated materials
with quarter filled bands. Thus we aim at understanding a whole class of materials on a common basis.
 
\section{Periodic distortions in interacting chains}
Charge order -- independent of magnetic ordering -- in less than half-filled bands 
is found, for example, in organic charge transfer salts,  
like the Bechgaard salts (TMTSF)$_2$X,  with X=PF$_6$, AsF$_6$, ClO$_4$, ReO$_4$, or Br, or
their sulfur analogs (TMTTF)$_2$X.  These
systems, which consist of stacks of organic molecules forming weakly coupled 
one-dimensional chains, exhibit a large variety of low temperature phases \cite{Dumm,Schwing01}.
They  have been studied using a purely electronic model, namely the extended Hubbard model
with next-nearest-neighbor interaction. Recent results have been  obtained by mean-field approaches 
\cite{Kobayashi}  and exact diagonalization \cite{Clay}. On the other hand, charge order is often
connected to a structural transition and hence to electron-phonon coupling.
The coupling of a one-dimensional metal to an elastic lattice results in an instability towards a
periodic lattice distortion commensurate to the band-filling, known as Peierls transition.
Within a lattice model, the Peierls transition is captured by a modulated 
hopping term  in the Hamiltonian of the non-interacting system \cite{SSH}.
A periodic potential, instead of a modulated hopping, has similar effects on the electronic structure
of the one-dimensional metal, leading to a band insulator. 

The interplay of electron-lattice  and electron-electron interaction is conveniently studied in the 
framework of the Hubbard model. With interaction, the Peierls and the so-called ionic model show
different ground-state phase diagrams \cite{Fabrizio}. Many results are available for the half-filled Hubbard
model with periodic hopping or potential,  for which the transition from a band insulator to a Mott insulator
has been studied in \cite{Maki,Kampf,Schoenhammer,Fehskea,Fehskeb}.
On the other hand, away from half filling no insulator-insulator-transition is present,
but a variety of ordering processes have to be taken into account \cite{Schmitteckert}.
The stability of the different ground states of the quarter-filled Hubbard model with respect to a periodic
potential has been investigated \cite{Penc,Yoshioka} less extensively as compared to the half-filled model.
The phase diagram of the quarter-filled Peierls-Hubbard model is discussed in \cite{Riera}. 
Hence, we restrict our calculations to the ionic potentials.
 
We access the stability regions
of the ionic Hubbard model by introducing  several potentials connected with  
different ordering processes of charge and spin degrees of freedom into the Hamiltonian.
In particular, we  investigate which potential leads to the largest energy gain, and 
analyze against which potential or  order the system is most unstable, indicating
which order will be established in nature.
In our calculations, we concentrate on the ionic Hubbard model at quarter filling, for both 
weak and  strong interaction. We determine the ground state and the ground state energy
as a function of interaction, doping and potential strength, using the density matrix renormalization group 
method \cite{dmrg}. Thus, our calculations give detailed insights into 
the properties of hole-doped antiferromagnets. 

\section{Models}
In the following section we first discuss the properties of
non-interacting electrons on a chain with different periodic potentials and the phase diagram of the
homogeneous Hubbard chain.  In the last subsection we discuss some basic features of the periodic Hubbard model.
\subsection{Peierls model}
The Peierls model is the prototypical model to study the coupling of non-interacting electrons to the 
lattice. The Hamiltonian is given by
    \begin{equation}
H_{\rm Peierls}=-\sum^{N}_{i,\sigma}
 t_i\left( c^+_{i,\sigma}c^{}_{i+1,\sigma} + {\rm h.~c.}\right)\; ,
    \end{equation}
    where we consider a modulated hopping term with $t_i=t[1+u\cos(Qa\cdot i)]$. $N$ denotes 
    the number of lattice sites and $a$ the lattice constant. In addition,
    $N_e$ is the number of electrons on the chain and $Q$ the wave vector of the periodic distortion.
    In case of a commensurate distortion we have $Q=2mk_F$, $m=1,2,\ldots$.
The dimerized chain, $Q=\pi/a$, at half filling, $n=N_e/N=1$, 
was investigated by Su, Schrieffer, and Heeger (SSH) \cite{SSH}. Due to the periodic distortion a band gap,
$\Delta=4ut$, opens at $k_\Delta=Q/2$. Since at half filling we have $k_\Delta=k_F$, the system is insulating. 
As is characteristic for a band insulator, the gaps for charge and spin excitations are identical.
In the following, we concentrate on the quarter-filled band and $4a$-periodic distortions, because
in case of the $2a$-periodic distortion the quarter-filled system is metallic, since $k_\Delta\neq k_F$.

To begin, we determine the energy bands and the energy gain due to a periodic modulation of the hopping
parameter with $Q=\pi/2a$. For quarter filling, $k_F=\pi/4a$, this system shows 
the Peierls transition, since $Q=\pi/2a=2k_F$. 
The Fourier transformation of the hopping term yields, in analogy with the SSH-model, the expression 
\begin{equation}
H_{\rm Peierls}=\sum_k
\left(c^+_k \  c^+_{k+Q} \ c^+_{k+2Q} \ c^+_{k+3Q}
\right)
t\hat{h}_{\alpha\beta}
\left( \begin{array}{c} c_k  \\ c_{k+Q} \\ c_{k+2Q} \\ c_{k+3Q}
\end{array}\right) \;,
\end{equation}
where the  matrix $\hat{h}_{\alpha\beta}$ is given by
\begin{widetext}
\begin{equation}
\hat{h}_{\alpha\beta}=\left(\begin{array}{cccc}
-2\cos{ka} &-u({\rm e}^{ika}+i{\rm e}^{-ika})  & & u({\rm e}^{ika}-i{\rm e}^{-ika}) \\
u(i{\rm e}^{ika}-{\rm e}^{-ika}) & -2\sin{ka}& -u({\rm e}^{-ika}+i{\rm e}^{ika})& \\
 & u(i{\rm e}^{-ika}-{\rm e}^{ika}) &2\cos{ka} & u({\rm e}^{ika}+i{\rm e}^{-ika})\\
  u({\rm e}^{-ika}+i{\rm e}^{ika})& &u({\rm e}^{-ika}-i{\rm e}^{ika}) &2\sin{ka}
  \end{array}\right).
  \end{equation}
  \end{widetext}
  The eigenvalues of this matrix are 
  \begin{equation}\label{eq58}
  \varepsilon(k) =\pm t\sqrt{2\left(1+2u^2\pm\sqrt{\cos^2{2ka}+4u^2(1+\sin^2{2k})}\right)},
  \end{equation}
  compare \cite{Hida}. The four energy bands are separated by the energy gaps
  \begin{equation}
  \Delta_1=4ut , \quad \Delta_2=2u^2t, \quad {\rm and} \quad \Delta_3=\Delta_1,  
  \end{equation}
  where $\Delta_1=\varepsilon_2(k=k_F)-\varepsilon_1(k=k_F)$ and 
  $\Delta_2=\varepsilon_3(k=2k_F)-\varepsilon_2(k=2k_F)$.
  The energy gain of the fermions, for the quarter-filled case, obtained by summing up the
  contributions of all occupied states, is found to be $\propto u^2t\ln{u}$, as for half filling.

\subsection{Ionic model}
\begin{figure}
{\includegraphics[width=0.45\textwidth]{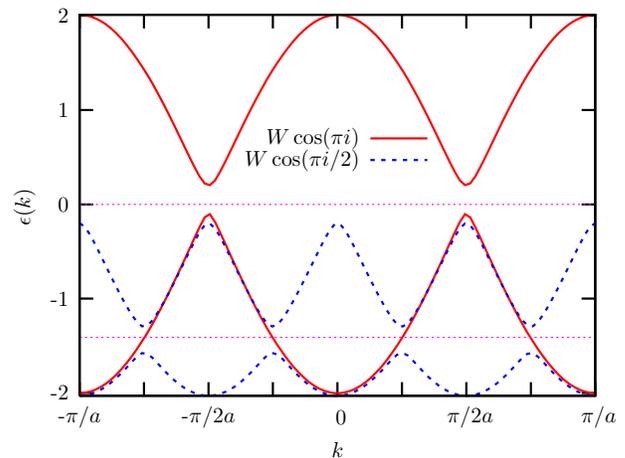}}
\caption{(Color online) One-particle energy $\varepsilon(k)$ versus momentum $k$, where $W=0.1t$ and the horizontal lines indicate the Fermi level of the half-filled and quarter-filled band.
The straight line corresponds to the $Q=\pi/a$-periodic potential, the dashed line to
the $Q=\pi/2a$-periodic potential. The reduction of the Brillouin zone of the 
clean model is obvious.} 
\label{fig-band}
\end{figure}

Another type of periodic distortion is given by local potentials, resulting in the  so-called ionic model
\begin{equation}
H_{\rm ionic}=-t\sum^{N}_{i,\sigma}
 \left( c^+_{i,\sigma}c^{}_{i+1,\sigma} + {\rm h.~c.}\right)
 +\sum^{N}_{i}W_i n_i
    - \sum^{N}_{i} h_iS_i^z\;.
  \end{equation}
Again, for the half-filled band the same calculation as before can be performed for a periodic 
potential, with $\Delta_c=\Delta_s=2W$. In case of the ionic model for a quarter-filled system 
we compare several potentials, which correspond to different charge and spin patterns: 
\begin{itemize}
\item[a)] First, we study a simple potential with 
$2k_F$($4a$)-period, given by $ W_i=W\cos(\pi i/2)$. Here, the charge order for zero hopping
is given by 1/0.5/0/0.5 electrons per site, periodically continued. Spin order is not established.
\item[b)] A modification of a) given by  $W_i=W\cos(\pi i/2)+W\cos(\pi (i+1)/2)$, which
leads to  two occupied and two unoccupied sites (``cluster") \cite{Clay}, i.~e.\ a 1/1/0/0 charge order.
\item[c)] A charge order with a 1/0/1/0 pattern, i.~e.\ a $4k_F$($2a$)-periodic order,
is forced by a potential with $W_i=W\cos(\pi i) $. With interaction we likewise expect  magnetic ordering.
\item[d)]In addition, a  local magnetic field with $h_i=W\cos(\pi i/2)(n_{i,\uparrow}-n_{i,\downarrow})$ 
enforces a $\uparrow$/0/$\downarrow$/0 pattern. As in c), we expect a 2$k_F$($4a$)-period  
for the spins and correspondingly a $4k_F$($2a$)-period for the charges. 
\end{itemize}
Diagonalizing the resulting Hamilton matrix, as described above, yields the  dispersions
\begin{itemize}
 \item[a),] d) \quad
 $\varepsilon(k)=\pm t\sqrt{2+w^2/2\pm\sqrt{4\cos^2(2k)+2w^2+w^4/4}}$,  
 \item[b)]  $\varepsilon(k)=\pm t\sqrt{2+w^2/2\pm\sqrt{4\cos^2(2k)+4w^2}}$,
    \item[c)]  $\varepsilon(k)=\pm t\sqrt{w^2+4\cos^2k}$, 
    \end{itemize}
    with $w=W/t$.
Fig. \ref{fig-band} shows the dispersion for  potentials  a) and c).
In comparison to the Peierls model not only a gap opens  at $k=Q/2$ but the lower and upper bands 
are shifted down and up, respectively. In case of a small perturbation, the dispersion is  similar for the
Peierls and the ionic model, as well as for potentials a) and b).
Potential a) and d) are equivalent in the non-interacting case.

The energy gain in the quarter filled band is found to be about  $-W^{1.65}$ in cases a), b), and  d).
For small $W$, the energy gain here is nearly the same in case a) and b), whereas it becomes weaker in
case b) than in case a) at $W\approx 0.5t$. In case c), the energy gain is quadratic,
$E(W)-E(0)=\frac{8}{3}W^2$. It mainly traces back  to the band shift.

Accordingly, the gaps are given by
\begin{itemize}
 \item[a),] d) \quad
$\Delta_1=W$, $\Delta_2=\sqrt{2}W$, $\Delta_3=W$;
   \item[b)] $\Delta_1\sim W$ for small $W$, $\Delta_1 \to \Delta_1^\infty$
	    for strong $W$, $\Delta_2=\sqrt{2}-W/2$, $\Delta_3=\Delta_1$;
  \item[c)] $\Delta_1=0$, $\Delta_2=W$, $\Delta_3=0$.
		    \end{itemize}
\subsection{Hubbard model}
The Hubbard model is known to capture the interplay between kinetic energy (delocalization) 
and interaction (localization) in electronic systems. The Hamiltonian is given by
\begin{equation}
H_{\rm Hubb}= -\sum^{N}_{i,\sigma}
 t_i\left( c^+_{i,\sigma}c^{}_{i+1,\sigma} + {\rm h.~c.}\right)
    + U\sum^{N}_{i}n_{i,\uparrow}n_{i,\downarrow} .
    \end{equation}
The Hubbard model in one dimension is exactly solvable by means of the Bethe ansatz \cite{Lieb}.
Note that in one dimension another  useful  
formulation of the Hubbard model is available on the basis of the bosonization technique.
The low lying excitations of the non-interacting as well as the
interacting fermions system are sound waves, i.~e.\ the Fermi system can be described
as a non-interacting Bose system,  called a Luttinger liquid,  showing 
spin-charge separation. In the clean case, the Hubbard model has three phases. For $U < 0$, the
spin excitation spectrum has a gap  and the low-lying charge
excitations can be described by those of a Luttinger liquid. For $U> 0$ and away from half filling,
spin and charge excitations are those of a Luttinger liquid.
The last phase occurs for $U > 0$ and half filling, where the charge excitations have a
gap and the spin excitations are of Luttinger type.
A relevant $4k_F$-Umklapp scattering term, only present for half filling,
is responsible for the Mott gap in the latter phase.
The bosonization technique is adequate for metallic systems or in the weak coupling regime.
It is useful to determine the phase boundary between metals and insulators but it is not suitable
for distinguishing different insulating phases for intermediate or strong perturbations. 

\subsection{Periodic Hubbard model}
A commensurate periodic distortion -- i.~e.\ commensurate to the band filling -- 
introduces an additional non-linear term in the bosonized Hamiltonian, which
couples spin and charge degrees of freedom and destroys the integrability of the clean Hubbard model.
In the half-filled case we therefore find a transition between two possible insulating phases, i.~e.\ 
the band (Peierls) insulator ($W, ut \gg U$) with charge and spin gap of the same order and the Mott
insulator ($ W, ut \ll U$) with a charge gap proportional to the interaction, but vanishing spin gap, where
periodic distortion and interaction oppose each other. However, the transition is fundamentally different
for the Peierls and the ionic Hubbard model \cite{Fabrizio,Torio01}. 
In the Peierls model the transition between the two phases is smooth and continuous \cite{Mocanu}, whereas
a third phase -- a spontaneous dimerized insulating phase -- is found in the ionic Hubbard model.
Within this phase, charge and spin gap are non-zero, and the dimerization operator has a non-zero
expectation value. In fact, the extension of this  phase is discussed controversially.

The situation in the quarter filled band is completely different. With interaction, no insulating phase is
established, except for very strong interaction, where higher order terms of the Umklapp scattering
can cause a Mott insulator on its own \cite{Yoshioka}.
In the quarter-filled case with $2k_F$-periodic potential we expect from bosonization
a transition from the Luttinger liquid to the band insulator in the non-interacting as well as in the
interacting system. The interaction modifies the exponents
of the gaps and the energy gain. Since they become smaller with increasing interaction, 
interaction stabilizes the band insulator.
Here, the interplay between periodic distortion and interaction is weak, since the quarter filled model
with interaction is still a metal. A similar situation is found for half filled spinless fermions for weak
interaction, where the Umklapp scattering
induces the insulator at finite interaction strength \cite{Schuster98}.
In case of a $4k_F$-periodic potential Umklapp scattering  becomes relevant \cite{Penc}
due to the doubling of the unit cell or, equivalently, due to the reduction of the Brillouin zone, see 
the straight line in Fig.\ 1.
In this case, the system undergoes a transition from the Luttinger liquid to a Mott insulating phase
when the potential and the interaction are present. Potential and interaction hence cooperate rather than
oppose each other as in the half filled ionic model.

An analysis of the ionic Hubbard model with another $2k_F$-periodic potential
is given in \cite{Torio06}. The authors find the band insulator for weak
interaction but strong potential. Our $4k_F$-periodic pattern, hence the Mott insulator, is obtained 
in their model in the limit of strong interaction. Both phases are separated by a bond ordered phase.
Thus, the model potential used in \cite{Torio06} reveals the features of the half filled ionic Hubbard 
model. The analysis, however, concentrates on large $|U-W|$.
In our case, the $4k_F$-periodic pattern is not a limit of the $2k_F$-periodic patterns.

In the following, we concentrate  on the ionic Hubbard model at weak to intermediate potential strength
and intermediate to strong interaction. In addition, we study the region of stability for each potential.
\section{Numerical results for the ionic Hubbard model at quarter filling}
In the following we discuss the numerical data for the ground state energy, 
the charge and spin gap, and the spin-spin correlation function.
The numerical results are obtained by  the density matrix renormalization group method (DMRG), as implemented
by Brune \cite{BruneDiss}.
The DMRG is a quasi-exact numerical method to determine the ground state
properties, i.~e.\ the ground state and the ground state energy, of long one-dimensional 
(non-integrable) systems with reasonable accuracy \cite{dmrg}. 
Regarding different boundary conditions, it is useful to take into account the equivalence of fermion
and spin models and to implement the spinless-fermion model in terms of an equivalent spin
chain, and the Hubbard model as two coupled spin chains without perpendicular XY-coupling.
Using the DMRG, it is possible to extend the tractable system lengths for the Hubbard chain to 
$N=60\ldots100$ sites. In our simulations we perform five lattice sweeps and keep 300
to 500 states per block. Correlation functions can be obtained
within an error of 10$^{-6}$ in the Hubbard model when using open boundary
conditions. A memory of about 700 MB is required.
\subsection{Energy}
\begin{figure}
\includegraphics[width=0.5\textwidth]{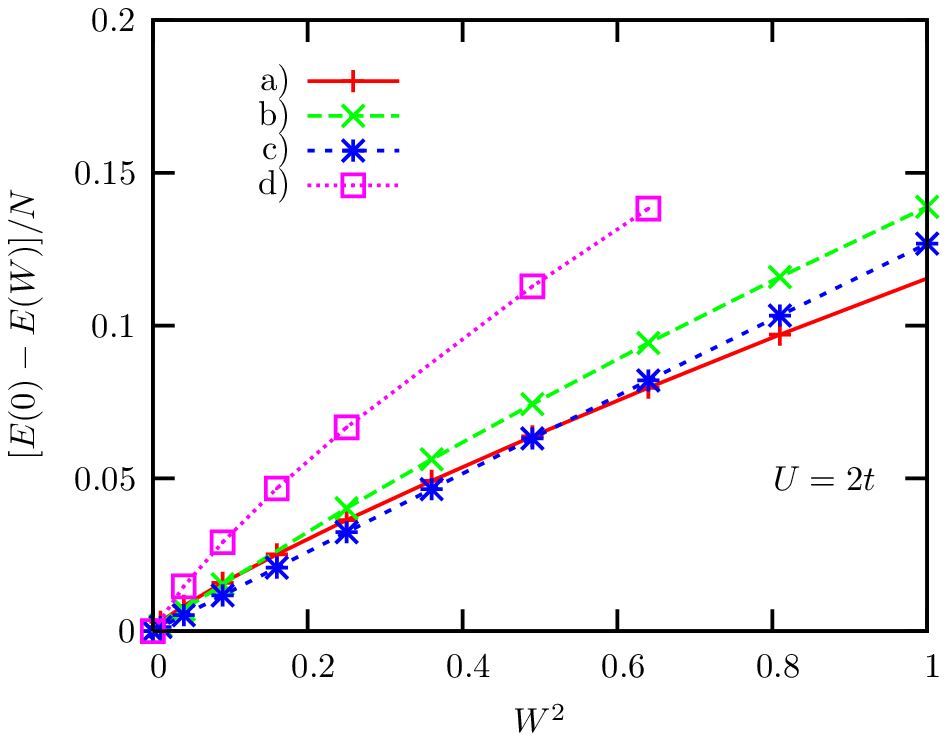}\\[0.5cm]
\includegraphics[width=0.5\textwidth]{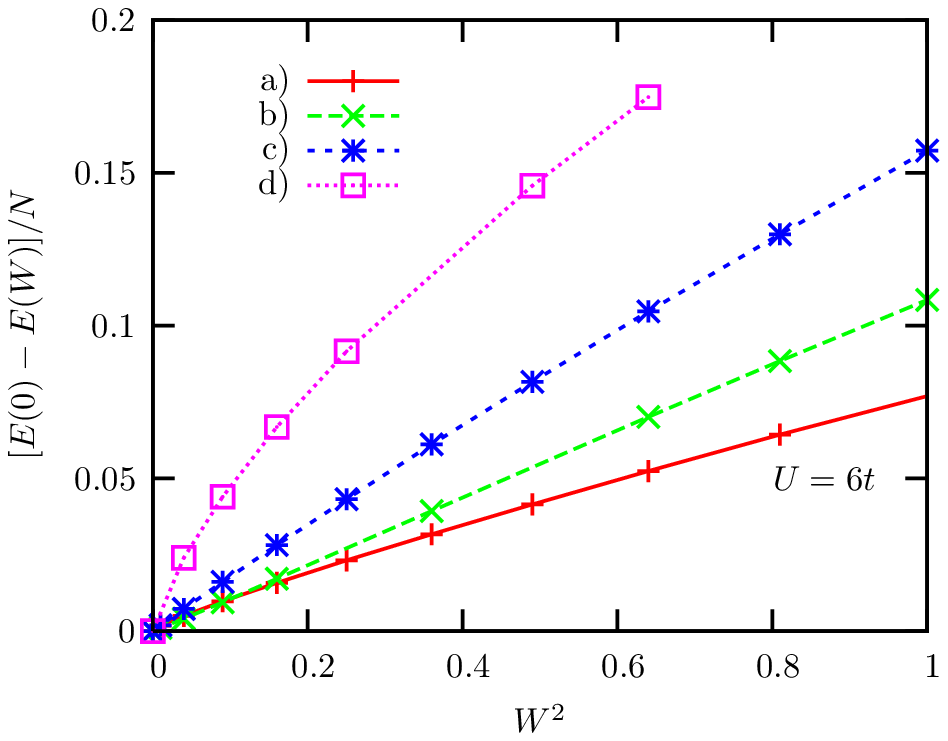} 
\caption{(Color online) Energy gain per site versus square of the potential strength (both in units of $t$).
We compare potentials a), b), c), and d). The number of sites varies between $N=28$ and 
$N=60$, where $[E(0)-E(W)]/N$ is almost independent of $N$. The interaction is $U=2t$
in the upper plot and $U=6t$ in the lower plot.} \label{fig EN} \end{figure}

First, we determine the ground state energy as a function of $W$ and $U$. 
The energy gain depends almost quadratically on the potential strength for all potential types, and is
dominated by the band shift. The numerical data are depicted in Fig.\ \ref{fig EN}.
In case of a coupling to the charge density, $W_in_i$,
the 2$k_F$-periodic potential is stabilized for small $U$ (case a)), 
and the 4$k_F$-periodic potential (case c)) for large $U$, as found for the Hubbard-Holstein model \cite{Riera99}.
This energy gain is mainly due to the  prefactors. 
A more detailed analysis of the algebraic behavior for $W\to 0$ shows
that $E_a\propto W^{1.68}$ and $E_c\propto W^2$ if $U=2t$, but
$E_a\propto W^{1.76}$ and $E_c\propto W^{1.76}$ if $U=10t$. In  case a) the exponent increases with 
interaction, in  case c) it decreases.
For all interactions, however, term d), $h_iS^z_i$, is dominant.
Of course, in a real system, a potential coupled to the density is much bigger than a magnetic field.
It leads -- like potential c) -- to 4$k_F$-oscillations in the charge density and 2$k_F$-oscillations 
in the magnetization. In the non-interacting system the data points for potentials a) and d) lie
nearly on top of each  other (by  use of open boundary conditions the exact equivalence of a) 
and d) is waived), but in the interacting system the energy gain due to potential d) grows faster  
with interaction than in all other cases. To be more specific, we find $E_d\propto W^{1.62}$ for
$U=2t$ and $E_d\propto W^{1.44}$ for $U=6t$.
\subsection{Charge and spin gap}
\begin{figure}
\includegraphics[width=0.5\textwidth]{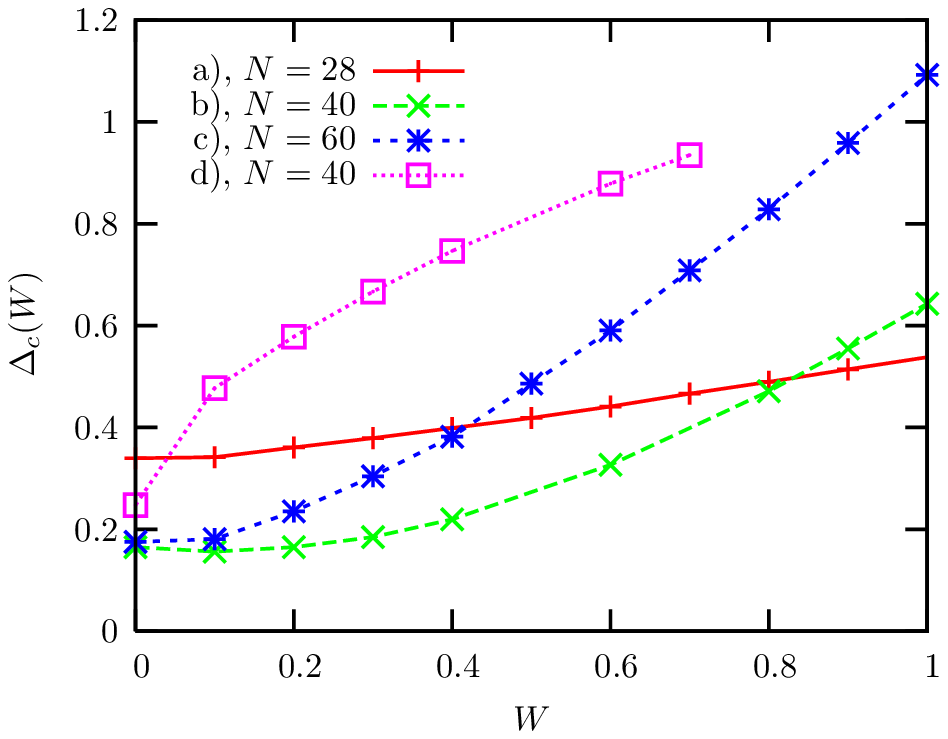}\\[0.5cm]
\includegraphics[width=0.5\textwidth]{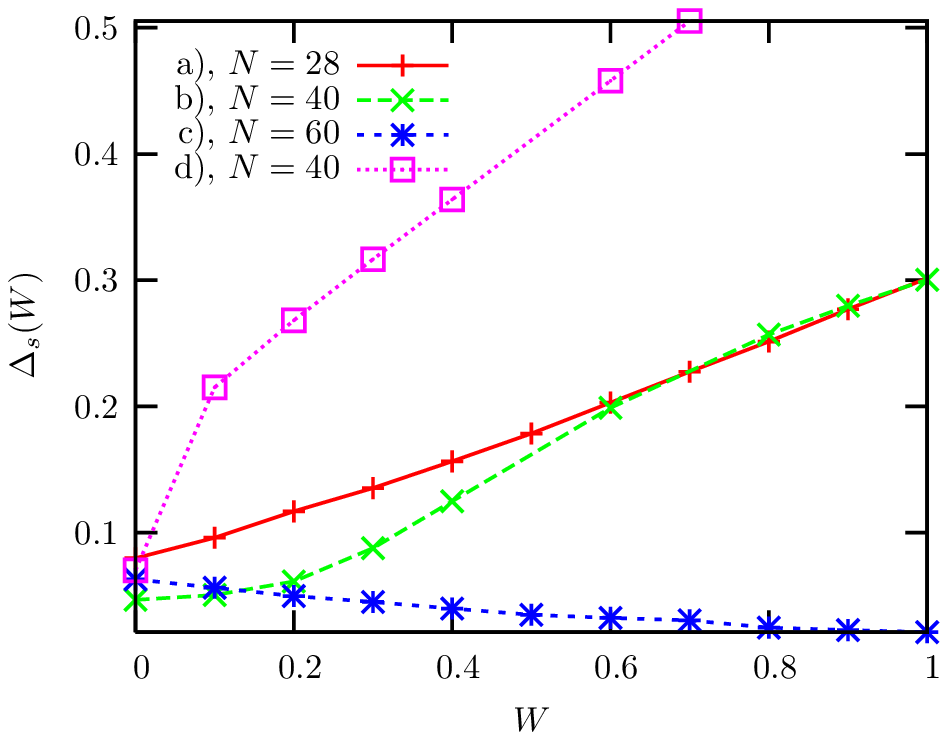}
\caption{(Color online) Charge  and spin gap versus potential strength (both in units of $t$). In the upper
plot we show $\Delta_c$ and in the lower $\Delta_s$, comparing potentials a), b), c), and d).
The number of sites varies between $N=28$ and $N=60$ and the interaction strength is $U=6t$.}
\label{fig-gap}
\end {figure}

We note that the excitation gaps $$\Delta_c=\frac{1}{2}[E(N_e+1)+E(N_e-1)-2E(N_e)]$$ and 
$$\Delta_s=\frac{1}{2}[E(S^z=1)+E(S^z=-1)-2E(S^z=0)]$$ are calculated for finite systems only. 
An extrapolation to the thermodynamic limit, $\Delta=\Delta_0+f(N)$ with $f(N)\to 0 $ for $N\to \infty$, 
is not performed, for the following reasons: 
for small gap $\Delta_0$ the accuracy of the DMRG away from half filling is not accurate enough  
and for large  $\Delta_0$ we have $f(N)\approx 0$ for the considered system sizes.
The dependence of the excitation gaps on the potential strength is shown in Fig. \ref{fig-gap}. 
In cases a), b) and d) we see that $\Delta_s\sim \Delta_c$, according to the band insulating state.
An almost linear dependence on $W$, as for non-interacting particles,
is recovered in the interacting system for the charge gap in cases a), b), and c).
The saturation of the gaps in case b), as discussed in Sec. 3B, is obtained for
the spin gap in the considered parameter regime. 
The influence of the interaction in the band insulating state can be summarized as follows.
In case a) the charge gap becomes smaller with interaction, the spin gap even more.
In case b) the charge gap increases with  interaction for small $U$, but saturates for strong interaction.
The spin gap decreases rapidly with interaction.
In case d) the charge gap increases but does not  saturate. The spin gap shows a maximum for intermediate
interaction strength. For $U=10t$ it is even smaller than in the non-interacting system. On the contrary,
we find for potential c) a linear increase of $\Delta_c$ with $U$ and $W$ 
as well as $\Delta_s(W, U, N) \sim 1/U, 1/W, 1/N\to 0$, indicating the Mott insulating state.
Due to the finite system size, we are not able to  obtain exponents for small $W$. As a consequence,
a comparison with bosonization or  the energy data, see Section A, is not possible.
To conclude, for strong interaction and small potential we find a large charge gap but a small spin gap.
\subsection{Correlation Function}
In order to calculate the spin-spin correlation function $\langle S^z_iS^z_j\rangle$ within DMRG we again
use open boundary conditions and the Parzen filter function \cite{Kuehner} to reduce the Friedel oscillations.
In the Mott insulating regime, the Friedel oscillations in the spin sector are long ranged, but decay fast in
spin gap systems \cite{Cs04}. The main question concerns the correspondence of the spin-spin correlation
functions of the $2a$-periodic potential (case c))
and the Heisenberg model. For comparison, we calculate the
spin-spin correlation function of the half-filled Hubbard model, which can be mapped onto a
Heisenberg chain in case of strong interaction. In the Heisenberg model we have
$$\langle S^z_iS^z_0\rangle = \frac{1}{4\pi (a\cdot i)^2} +\frac{a_{2k_F}\cos(2k_Fa\cdot i)}{a\cdot i}.$$ 

In gapped systems, the spin-spin correlation functions reflect the behavior of
the energy gaps. Thus, the spin-spin correlation function decreases
exponentially in cases a) and b), which show a spin gap without spin order. 
In case d) the spins are fixed by the potential $h_i$. 
The exponentially decaying part is hard to extract from the mean magnetization in the numerical data due to the 
incomplete suppression of the boundary oscillations. 

\begin{figure}
\includegraphics[width=0.5\textwidth]{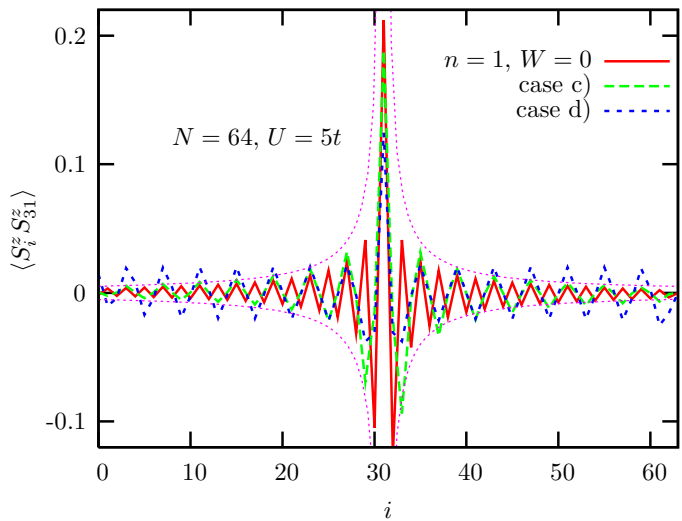}
\caption{(Color online) Spin-spin correlation function $\langle S^z_iS^z_{\rm 31}\rangle$ versus site $i$. 
We compare the half-filled Hubbard chain (straight line; 
which is equivalent to a Heisenberg chain), with the
quarter-filled Hubbard chain on a $4k_F$-periodic potential (long-dashed line, $W=t$;  
equivalent to a half-filled Heisenberg chain), and the quarter-filled Hubbard chain in a $2k_F$-periodic 
magnetic field (short-dashed line, $W=0.1t$; equivalent 
to a half-filled Ising chain). In all data sets the interaction is $U=5t$.} 
\label{fig-corr}
\end{figure}

In the Mott insulating regime, however, the linear decrease of the spin-spin correlation function with distance
is obtained already for  small potential $W$. For  better comparison with the data of the half-filled Hubbard
chain, we show  data for $W=t$ in Fig. \ref{fig-corr}. Obviously, the quarter-filled Hubbard chain with
4k$_F$-periodic distortion can be mapped onto the Heisenberg model, too.
We note that in case d) already a weak potential -- connected with a small spin gap -- leads
to significant structures in $\langle S^z_iS^z_j\rangle$, whereas in case c) the potential 
has to be  much stronger to yield effects of similar size. The increase of $S(q)\propto q$
appears both in the half-filled Hubbard chain and in the quarter-filled Hubbard chain with $4k_F$-periodic
potential. $S(q)$ has a sharp maximum at $q=\pi/2$ in case d). 
\section{Summary}
In summary, we have studied the effects of periodic potentials on a chain. In particular, 
we have considered the one-dimensional Hubbard model at quarter filling. 
We have compared periodic potentials yielding different types of behavior:
a) a 2$k_F$-periodic potential leading to a band insulator,
b) a 2$k_F$-periodic potential leading to a band insulator with cluster-like arrangements of the charges,
c) a 4$k_F$-periodic potential leading to a Mott insulator with antiferromagnetic alignment of
Heisenberg type spins on next-nearest neighbor sites, 
and d) a 2$k_F$-periodic magnetic field leading to a band insulator with antiferromagnetic alignment of Ising
type spins. Cases c) and d) reveal the same charge and spin distribution but different spin excitations.
In the Heisenberg case the spin excitations are gapless, whereas in the Ising case they are gapped.

In non-interacting systems, we find $E_a=E_d>E_b>E_c$, thus the band insulator.
The charge distribution is quite homogeneous, and the cluster like arrangement is not minimal in energy.
Turning on the interaction, potential d) results in the largest energy gain, while the order of the remaining
potentials depends on both the potential strength and the interaction. 
The interaction strongly supports the spin order, where double occupancy is suppressed.

In the following discussion we rely on the potentials a) to c), where the spin-order, if present, is a result
of the charge order. For weak interaction and weak potential, we recover the behavior of the non-interacting
system, with $E_a>E_b>E_c$. Turning to intermediate values ($U\approx 3t$, $W\approx t$ ), 
the cluster-like arrangement of potential b) 
gains more energy than the homogeneous distribution of case a), $E_b>E_a>E_c$.
For strong $W\to t$, also a 4$k_F$ pattern is favored against the homogeneous case, $E_b>E_c>E_a$. 
Thus, for small or intermediate interaction and potential,
the gain of energy due to the hopping is stronger than the 
effects of the repulsive interaction or the underlying potential. With increasing potential, however,
the effects of the potential dominates.

On the other hand, regarding a weak potential but increasing interaction,
we find that a pattern related to potential a) is established only at small interaction. The
cluster-like arrangement is formed
at intermediate interaction ($U\approx 3-5t$), but at strong interaction
the $4k_F$-periodic pattern dominates. Concerning the  
spin chain compound (Ca,Sr)$_{14}$Cu$_{24}$O$_{41}$, we have 
applied the potential only in order to clarify the leading instability of the system.
For comparison with a real material which shows no lattice effects, 
the limit $W\to 0$ is relevant. In addition,
for copper oxides usually a correlation parameter of $U\approx 8 eV$ \cite{Anisimov} is assumed.
Together with the band width of about 1 eV, see Fig. 1, we have $U > 15t$.
In this parameter region, we find -- neglecting the magnetic field -- 
the 4$k_F$-periodic charge pattern with antiferromagnetism to be the most probable ground state.
The energetic order  found for weak potentials extends to strong potentials.
For strong interaction, we always have $E_c>E_b>E_a$.
The tendency of the repulsive interaction to separate the charges dominates the phase diagram 
in this region. The influence of the potential is weaker, leading to a preference of the cluster
over the homogeneous distribution.

Finally, we remark that adding electrons stabilizes the cluster formation.
This observation likewise agrees with  experimental results, obtained for the 
(Ca,La)$_{14}$Cu$_{24}$O$_{41}$ series, where the cluster size grows on additional charge.
Therefore we conclude, that the antiferromagnetism is due to the band filling, thus a commensurability effect. 

\subsection*{Acknowledgement}
We thank U.\ Eckern for helpful discussions, and P. Brune for providing the DMRG code.
Financial support by the Deutsche Forschungsgemeinschaft within
SPP 1073 and SFB 484 is acknowledged.

\end{document}